\documentclass{rstransa}%%%%where rstrans

\usepackage[acronym]{glossaries}
\usepackage[color=yellow]{todonotes}
\usepackage[outercaption]{sidecap}  
\usepackage[numbers]{natbib}
\usepackage{ulem} 

\usepackage{bm}
\newacronym{RIXS}{RIXS}{resonant inelastic X-ray scattering}
\newacronym{REXS}{REXS}{resonant elastic X-ray scattering}
\newacronym{tr}{tr}{time-resolved}
\newacronym{XAS}{XAS}{X-ray absorption spectroscopy}
\newacronym{FEL}{FEL}{free electron laser}
\newacronym{XFEL}{XFEL}{X-ray free electron laser}
\newacronym{LCLS}{LCLS}{Linac Coherent Light source}
\newacronym{FLASH}{FLASH}{Free-electron LASer}
\newacronym{FERMI}{FERMI}{Free Electron laser Radiation for Multidisciplinary Investigations}
\newacronym{SACLA}{SACLA}{SPring-8 Angstrom Compact free electron LAser}
\newacronym{PAL-XFEL}{PAL-XFEL}{Pohang Accelerator Laboratory X-ray Free Electron Laser}
\newacronym{SwissFEL}{SwissFEL}{Switzerland's X-ray free-electron laser}
\newacronym{FT}{FT}{Fourier Transform}
\newacronym{IXS}{IXS}{Inelastic X-ray Scattering}
\newacronym{SASE}{SASE}{Self-Amplified Spontaneous Emission}
\newacronym{FWHM}{FWHM}{Full-Width at Half Maximum}

\begin{document}

\newcommand{\ket}[1]{\left|{#1}\right>}
\newcommand{\bra}[1]{\left<{#1}\right|}
\newcommand{\abs}[1]{\left|{#1}\right|}

%%%%%%%%%%%%%%%%
%\todo[inline]{Try to address John's broader comments better in the review process}
%%%%%%%%%%%%%%%%%%%%
\title{Ultrafast dynamics of spin and orbital correlations in quantum materials: an energy- and momentum-resolved perspective}

\author{%%%% Author details
Y. Cao$^{1, 2}$, D. G. Mazzone$^{3}$, D. Meyers$^{1}$, J. P. Hill$^{3}$, X. Liu$^{4}$, S. Wall$^{5}$, M. P. M. Dean$^{1}$
}

%%%%%%%%% Insert author address here
\address{
$^1$Condensed Matter Physics and Materials Science Department, Brookhaven National Laboratory, Upton, New York 11973, USA \\
$^2$Materials Science Division, Argonne National Laboratory, Argonne, Illinois 60439, USA\\
$^3$National Synchrotron Light Source II, Brookhaven National Laboratory, Upton, New York 11973, USA \\
$^4$School of Physical Science and Technology, ShanghaiTech University, Shanghai 201210, China 
$^5$ICFO-Institut de Ci{\`e}ncies Fot{\`o}niques, The Barcelona Institute of Science and Technology, 08860 Castelldefels (Barcelona), Spain
}

%%%% Subject entries to be placed here %%%%
\subject{Condensed Matter Physics}

%%%% Keyword entries to be placed here %%%%

%%%% Abstract text to be placed here %%%%%%%%%%%%
\begin{abstract}
Many remarkable properties of quantum materials emerge from states with intricate coupling between the charge, spin and orbital degrees of freedom. Ultrafast photo-excitations of these materials hold great promise for understanding and controlling the properties of these states.  Here we introduce time-resolved resonant inelastic X-ray scattering (tr-RIXS) as a means of measuring the charge, spin and orbital excitations out of equilibrium. These excitations encode the correlations and interactions that determine the detailed properties of the states generated. After outlining the basic principles and instrumentation of tr-RIXS, we review our first observations of transient antiferromagnetic correlations in quasi two dimensions in a photo-excited Mott insulator and present possible future routes of this fast-developing technique. The increasing number of X-ray free electron laser facilities not only enables tackling long-standing fundamental scientific problems, but also promises to unleash novel inelastic X-ray scattering spectroscopies.
\end{abstract}

%%%%%%%%%%%%%%%%%%%%%%%%%%%
\maketitle
%%%%%%%%%% Insert the texts which can accommodate on first page in the tag "fmtext" %%%%%

\section{Introduction\label{sec:introduction}}
Quantum materials are a class of solids in which quantum mechanical effects are especially apparent generating states such as high-temperature superconductivity, colossal magneto-resistance, exotic magnetism and metal-to-insulator transitions, and are also amongst the least understood. One key observation is that many of these exotic phases emerge when antiferromagnetic parent compounds are chemically doped, revealing the rich physics of Mott insulators \cite{Lee2006Doping}. In these phases, the charge, spin, orbital and lattice degrees of degrees are intrinsically intertwined. This opens the fascinating prospect of photo-exciting one degree of freedom in order to modify another with the aims of generating new phases, improving the properties of known phases or disentangling the interdependencies among these degrees of freedom in order to better understand the ground state \cite{Zhang2014dynamics, Basov2017towards, Buzzi2018probing}. Probing the properties of these photo-excited states can, however, be challenging. In this article, we argue that the \gls*{tr}-\gls*{RIXS} technique has great potential for improving our understanding of ultrafast transient states.

\begin{SCfigure}
\includegraphics[width=8cm]{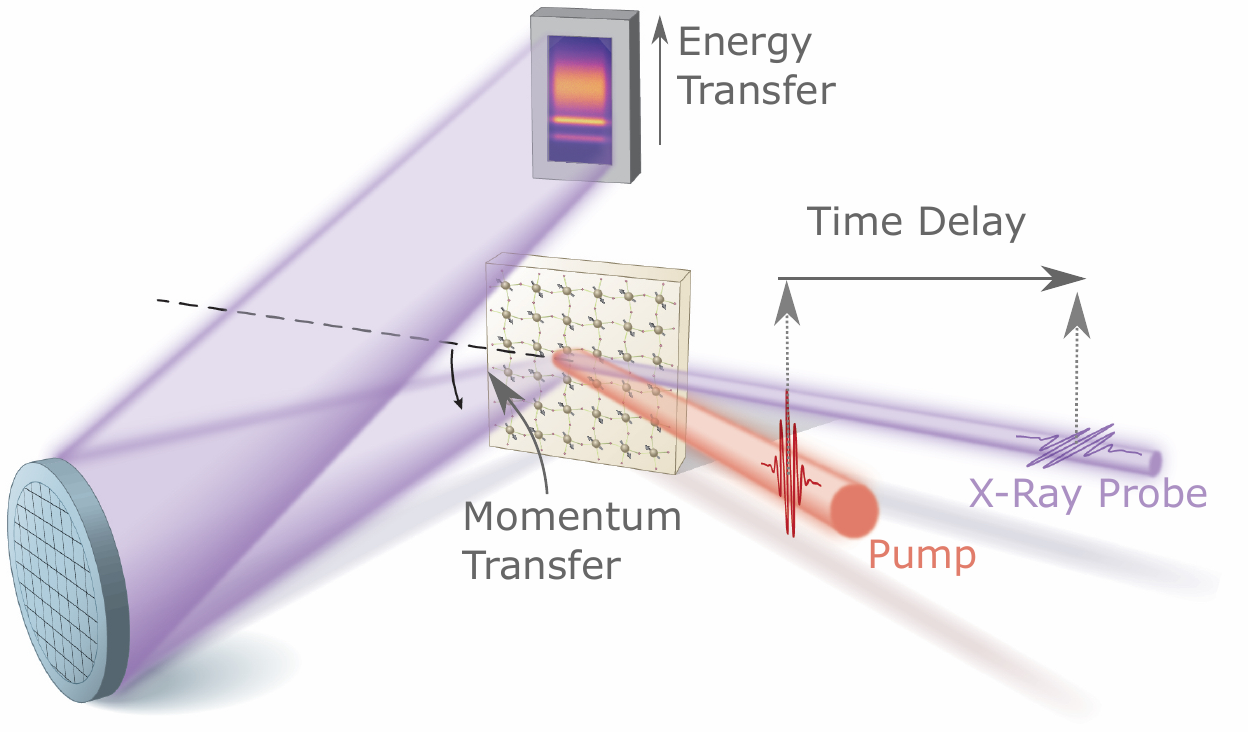}
\caption{Schematics of a tr-RIXS experiment. Generally in a RIXS measurement, the momentum transfer is determined by the angle between the incident and scattered X-rays from the sample, along with the wavelength of the X-ray. The energy of the scattered photons is resolved by a grating or crystal analyzer onto a CCD detector. In tr-RIXS a second pump laser pulse is used which arrives at a controllable time ahead of the incident X-rays. The tr-RIXS spectrum is then obtained by scanning the time delay.}
\label{fig:intro}
\end{SCfigure}

Figure~\ref{fig:intro} illustrates a typical \gls*{tr}-\gls*{RIXS} experiment. An optical pump pulse photo-excites the material into a transient state, which is then interrogated by an X-ray probe pulse. \gls*{RIXS} tracks the changes in energy and momentum of the scattered X-rays. Through energy and momentum conservation, the dispersion and lifetime of the spin, charge or orbital quasi-particles can be mapped throughout the Brillouin zone. This enables establishing minimal models and extracting the interaction parameters describing these quasi-particles. For example, the magnon energy at the Brillouin zone boundary in an antiferromagnet quantifies the strength of the magnetic exchange interaction.

Time-resolved \gls*{RIXS} offers new opportunities to investigate the dynamics of bosonic quasi-particles in the time domain. While facilities that can provide sufficiently short and bright X-ray pulses to carry out such experiments remain limited, the landscape is changing quickly with the increasing number of new X-ray sources over the next few years. The  article is organized as follows: in Sec.~\ref{sec:rixs} and Sec.~\ref{sec:instrumentation} we review the principles and instrumentation of tr-RIXS. We revisit our first measurements in a photo-excited Mott insulator Sr$_2$IrO$_4$ in Sec.~\ref{sec:results}, and discuss future technical developments that enable the search for novel physics in Sec.~\ref{sec:future}.

\section{RIXS as a Probe of Quantum Materials\label{sec:rixs}}
\gls*{RIXS} exploits energy and momentum transfer to infer a material's excitation spectrum. By tuning the incident X-ray energy to a core-hole resonance, it is possible to go beyond the typical selection rules of X-ray scattering and couple to magnetic, orbital and charge modes. Since X-rays have a wavelength comparable to the inter-atomic spacing in materials, the dispersion of these excitations can be measured across a sizable fraction of the Brillouin zone -- something that is not possible with infra-red or Raman spectroscopy. Such rich information, however, comes at a price as \gls*{RIXS} involves multiple interactions and can be challenging to interpret. Researchers often distinguish two types of processes: direct ``operator'' \gls*{RIXS} and indirect or ``shakeup'' RIXS \cite{ament2011resonant}. Direct \gls*{RIXS} involves excitations that are created by the photon absorption operators $\cal{D}$ and $\cal{D}'^\dagger$ described below. Indirect \gls*{RIXS}, on the other hand, arises due to the interaction of the core hole with electrons in the valence bands of the material. In cases where direct \gls*{RIXS} is not forbidden, such as the $L$ and $M$-edges of $d$ electron based transition metals, it tends to be the dominant process and this review will focus on direct \gls*{RIXS}. 

\begin{SCfigure}
\includegraphics{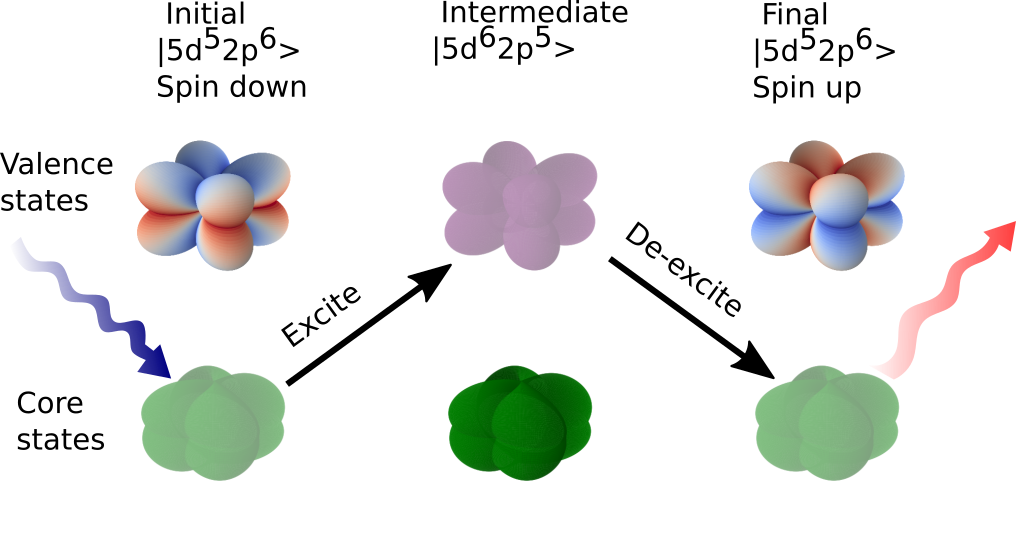}
\caption{An illustration of the $L_3$-edge RIXS process for a $5d^5$ iridate. Orbital states with (without) an active hole are in full color (translucent). Incident (scattered) X-ray are shown as blue (red) wavy arrows. The initial iso-spin down state is flipped to an up state via the spin-orbit coupling in the intermediate state.}\label{fig:rixs_process}
\end{SCfigure}

\gls*{RIXS} is formally described by the Kramers-Heisenberg formula, which involves transitions from the  ground state $\ket{g}$, through an intermediate state $\ket{n}$, to the final state $\ket{f}$ as (see Ref.~\cite{ament2011resonant})
\begin{eqnarray}
I({\bm Q}, \omega,{\bm \epsilon},{\bm \epsilon}^\prime)  \propto \left|   \sum_n\frac{\bra{f} {\cal D}'^\dagger  \ket{n} \bra{n} {\cal D} \ket{g}}{E_g+ \hbar\omega_{\bm k}-E_n+i\Gamma_n}\right|^2\delta(E_g - E_f + \hbar\omega).
\label{eq:K-H}
\end{eqnarray}
In this process, incident photons with wavevector ${\bm k}$, polarization ${\bm \epsilon}$ and energy $\omega_{\bm k}$ are scattered to states ${\bm k}^\prime$, ${\bm \epsilon}^\prime$ and $\omega_{{\bm k}^\prime}$ creating excitations with wavevector ${\bm Q} = {\bm k}^\prime - {\bm k}$ and energy $\omega = \omega_{{\bm k}^\prime} - \omega_{\bm k}$. $E_g$, $E_n$ and $E_i$ are the energies of $\ket{g}$, $\ket{n}$ and $\ket{f}$ and $\Gamma_n$ is the inverse core hole lifetime. Several important facts follow from a detailed consideration of Eq.~\ref{eq:K-H}. Since the X-ray wavelength ($1\sim 10$~\AA{}) is significantly larger than the extent of an atomic orbital ($\sim 0.1$~\AA{}), $\cal{D}$ and ${\cal D}'^\dagger$ can often be considered under the dipole approximation. As the dipole transitions depend on ${\bm \epsilon}$ and ${\bm \epsilon}^\prime$, the X-ray polarization can be exploited to help select the desired excitation \cite{van2006polarization}. The fact that two dipole operators are present means that, unlike in infra-red or Raman spectroscopies, electronic transitions within the same orbital manifold, often called $dd$-transitions, are allowed.  At an $L$-edge, $\ket{n}$ contains a strongly spin-orbit coupled $2p$ core hole, which means that the orbital angular momentum of the photon can be exchanged with the spin angular momentum in the valence band in order to create spin-flip excitations, such as magnons \cite{ament2009theoretical, braicovich2010magnetic}. This is illustrated schematically in Fig.~\ref{fig:rixs_process} and will be discussed in more detail in Sec.~\ref{sec:results}. One should note that \gls*{RIXS} is a coherent quantum mechanical process, and since $\ket{n}$ is not observed, the process involves a superposition of all possible intermediate states distributed throughout the lattice, which is why the collective, ${\bm Q}$-dependent excitations of the material are accessible. This is distinct from \gls*{XAS} where there is a core hole in the final state.\footnote{Here we use the term \gls*{RIXS} to include only processes without a core hole in the final state. Some papers in the literature use a broader definition.} Being a second-order process, \gls*{RIXS} has a relatively low cross-section compared to first-order processes, although this is offset somewhat by the enhancement of the scattering intensity on resonance \cite{ament2011resonant}.

\gls*{RIXS} is particularly well suited to probe magnetism in transient states. It is element and orbital resolved, bulk sensitive and compatible with small samples down to a few microns in size, which is important for pump-probe matching considerations. It is the only technique that has the proven ability to measure the dispersion of magnetic excitations in transient states. The best-established technique for studying these magnetic excitations, inelastic neutron scattering, is incompatible with ultra-fast experiments due to the slow velocity and large penetration depth of neutrons. Spin-polarized electron energy loss spectroscopy can also access magnetic dispersions \cite{vollmer2003spin}, but it has not yet been successfully implemented in an ultrafast experiment as space charge effects present significant challenges. 

Regardless of how it is measured, the full magnetic excitation spectrum can be formalized in the magnetic dynamic structure factor $S^{\alpha \beta}({\bm Q}, \omega)$, which characterizes the magnetic correlations in a magnetic material. This quantity is the space and time Fourier transform of the spin-spin correlation function $\left\langle s^{\alpha} s^{\beta}(r,t) \right\rangle $ between a reference spin component $s^{\alpha}$ and another spin $s^{\beta}(r,t)$ at distance $r$ and time $t$ $>$ 0. This fact makes it clear that similar information can, in principle at least, be obtained by either frequency domain or the time domain.  At excellent discussion of such correlation functions is provided in books such as Ref.~\cite{Chaikin1995principles}. It is further known that these correlation encodes the interactions present in the magnetic Hamiltonian \cite{Squires2012introduction}. In general, fast processes correspond to high frequencies and are often best probed in the frequency domain, whereas the converse is true to slow processes. 

Connecting $I({\bm Q}, \omega,{\bm \epsilon},{\bm \epsilon}^\prime)$ to $S^{\alpha \beta}({\bm Q}, \omega)$ is a topic of considerable active research \cite{ament2009theoretical,braicovich2010magnetic,letacon2011intense,dean2013persistence,jia2014persistent}. But the connection between these quantities is particularly well-established for Mott insulators based on $5d^5$ electrons, including iridates such as Sr$_2$IrO$_4$ \cite{kim2017resonant}. The interpretation of such spectral information will be covered in Sec.~\ref{sec:results}.

\section{Instrumentation\label{sec:instrumentation}}

The primary consideration of tr-\gls*{RIXS} instrumentations is to achieve sufficient time and energy resolution while ensuring reasonable photon throughput. \gls*{RIXS} is one of the most photon-hungry X-ray measurement techniques due to its small inelastic scattering cross-section. Recent advances in energy-analyzing spectrometer designs have made possible much-improved energy resolution and faster data collection at synchrotron light sources \cite{ghiringhelli2006saxes,dvorak2016towards, shvyd2013merix, kim2018quartz}. In this section, we will focus on aspects specific to the time-resolved experiments.

\subsubsection{Spatial overlap}

As is in most other pump-probe experiments, the probe X-ray photon footprint needs to be contained within that of the pump laser. Equally important, the penetration depth of the X-ray normal to the sample surface should not exceed that of the pump laser. At normal incidence, the X-ray penetration depth is usually much larger than that of the optical laser. To ensure a good match between the laser pumping volume and the X-ray probing volume, we could resort to one, or both of the following approaches - (1) using samples thinner than the penetration depth of the pump laser; and (2) aligning the X-ray to come in more grazing relative to the sample surface. Approach (1) could be achieved by thinning down bulk samples, or using films. For thicker samples, approach (2) needs to be used. The X-rays are usually at less than $2^\circ$ incidence angle relative to the sample surface and the laser is typically at e.g.\ $10^\circ$. This non-collinear geometry in approach (2) leads to decreased time resolution. Also, the large X-ray footprint due to glancing incidence gives rise to a large laser spot and hence reduced pump fluence.

The X-ray penetration depth in the soft X-ray regime ($\hbar\omega<2000$~eV) is usually a fraction of a micron on resonance. This makes it easier to implement approach (1) with a collinear pump-probe geometry. For harder X-rays ($\hbar\omega>4000$~eV), the penetration depth of a few microns requires approach (2) in many cases.

\subsubsection{Energy resolution}

The RIXS spectrometer works under the principle that energy dispersive optics distributes X-ray photons with different energies into slightly different directions. After prorating certain distance, X-ray photons with different energies are spatially separated, which can be differentiated by detectors with good spatial resolution. The following are some of the most important factors affecting the spectrometer energy resolution: (1) the length of the spectrometer arm; (2) the photon footprint along the energy dispersing direction; and (3) the size of the X-ray detector pixels. In the soft X-ray regime, ruled gratings are used to disperse X-rays, while high-quality diced single crystals are used for hard X-rays. A detailed description of the principles of spectrometer designs, as well as estimates of the energy resolution, are presented in the supplementary materials. State-of-the-art spectrometers often deliver sub-50~meV energy resolution \gls*{FWHM} relatively comfortably. Some of the more modern spectrometers and novel designs have an energy resolution approaching 10~meV \cite{dvorak2016towards,kim2018quartz}, though often only in a particular X-ray energy range.

With these advances in X-ray spectrometers, the incident X-ray needs to be well monochromated to achieve a high combined total energy resolution. The generic {X}-ray bandwidths from existing \gls*{XFEL}s are much larger than the resolution of the spectrometers. For example, in the \gls*{SASE} mode, the typical \gls*{FWHM} of the incident X-ray at 9~keV is $\sim$20~eV before entering the monochromator. Ideally the energy resolution of the X-ray shining on the sample should match that of the spectrometer. In this scheme, only a tiny fraction of the SASE beam (within much less than 100 meV) is expected to pass the monochromator for an optimized setup. To achieve a higher X-ray intensity at the sample without compromising the total energy resolution, we argue that \gls*{tr}-\gls*{RIXS} will benefit greatly from the seeded operation mode of \gls*{XFEL}s \cite{amann2012demonstration, Ratner2015experimental}. While the total X-ray flux exiting the undulator is reduced compared with that in the \gls*{SASE} mode, the incident photon flux within the desired energy bandpass will actually increase. For the current \gls*{XFEL}s, self-seeding tends to be more reliable in the hard X-ray regime.

As mentioned in Sec.~\ref{sec:rixs}, the different X-ray polarization dependences of different excitations can be used to help isolate the desired signal. Since magnetic X-ray scattering rotates the X-ray polarization, it is often useful to place the spectrometer as close as possible to $90^\circ$, using a horizontal scattering plane and horizontal incident X-ray polarization. In this way, the undesirable Thompson structural scattering is suppressed, and the visibility of the magnetic signal is enhanced. This works especially well for hard X-ray experiments, in which the whole Brillouin zone can be covered by moving the spectrometer only a few degrees.  With a horizontal scattering plane, the X-ray footprint will tend to elongate the beam horizontally, so setting the spectrometer to disperse the X-rays vertically is useful to maintain a constant vertical X-ray source size, independent of the scattering angle. 

\subsubsection{Momentum resolution\label{sec:momentum_resolution}}

An important advantage of \gls*{RIXS}, as compared with Raman spectroscopy, is the much larger momentum transfer between the incident and scattered photons. The main contribution to the momentum resolution is the acceptance angle of the \gls*{RIXS} spectrometer.  Typical momentum resolution is on the order of $\sim 0.01$~\AA$^{-1}$ for soft X-rays around 1~keV and $\sim 0.1$~\AA$^{-1}$ for hard X-rays around 10~keV. The worse momentum resolution in the hard X-ray regime is simply due to the larger photon momentum. In most cases the momentum resolution can be improved with a smaller acceptance angle provided there are enough scattered X-ray photons to make the experiment feasible. Notably, hard X-rays have significantly larger momentum transfer than soft X-rays, which allows access to multiple Brillouin zones even within the constraint of grazing incidence geometry needed for the penetration depth correction noted above.

\subsubsection{Time resolution\label{sec:time_resolution}}

In \gls*{tr}-\gls*{RIXS}, the temporal and energy resolution are fundamentally limited by the time-bandwidth product, $\Delta E \Delta \tau \sim \hbar$ setting an ultimate limit on the best resolution.  For a phase coherent Gaussian pulse with 100~meV energy resolution, the temporal resolution limit is approximately 40~fs. On top of the energy-time indeterminacy, the total time resolution has contributions from both, the durations of the X-ray pulse and of the pump optical laser, similar to other pump-probe experiments at \gls*{XFEL}s.

Another potential contribution to the time resolution can come from what is called the ``wave-front-tilt'' effect \cite{MOP019}. In this process the incident X-ray pulse entering the monochromator is stretched in time upon exit due to different X-ray photon path lengths through the optics. This effect tends to be more severe for longer-wavelength X-rays and for higher-resolution monochomators, and could reach a few picoseconds in the very soft X-ray regime with a single grating.    

The total time resolution can be further limited by the so-called ``jitter'', which is the uncertainty in the relative arrival time between the pump and probe pulses. The ``jitter'' tends to be between 20-100~fs depending on the \gls*{XFEL} in question and can be corrected for using a ``timing tool'' \cite{harmand2013achieving, kang2017hard}. There is also a potential geometrical contribution to the total time resolution arising from the relative optical path length difference between the X-ray and the pump laser over the photon footprint. The collinear geometry where the laser is parallel to the X-ray minimizes the geometrical contribution. This geometry is widely adopted in the soft X-ray elastic scattering experiments, and is expected to be used in \gls*{tr}-\gls*{RIXS} in the soft X-ray regime. For hard X-rays and in the presence of bulk samples, matching the pump-probe volume may require a substantial angle between the X-ray and the pump laser at the price of time resolution. One can, in principle, compensate for this effect by imparting a spatially dependent delay in the laser pulse. 

\section{First Results\label{sec:results}}
\begin{figure*}[tbh]
\includegraphics[width=\linewidth]{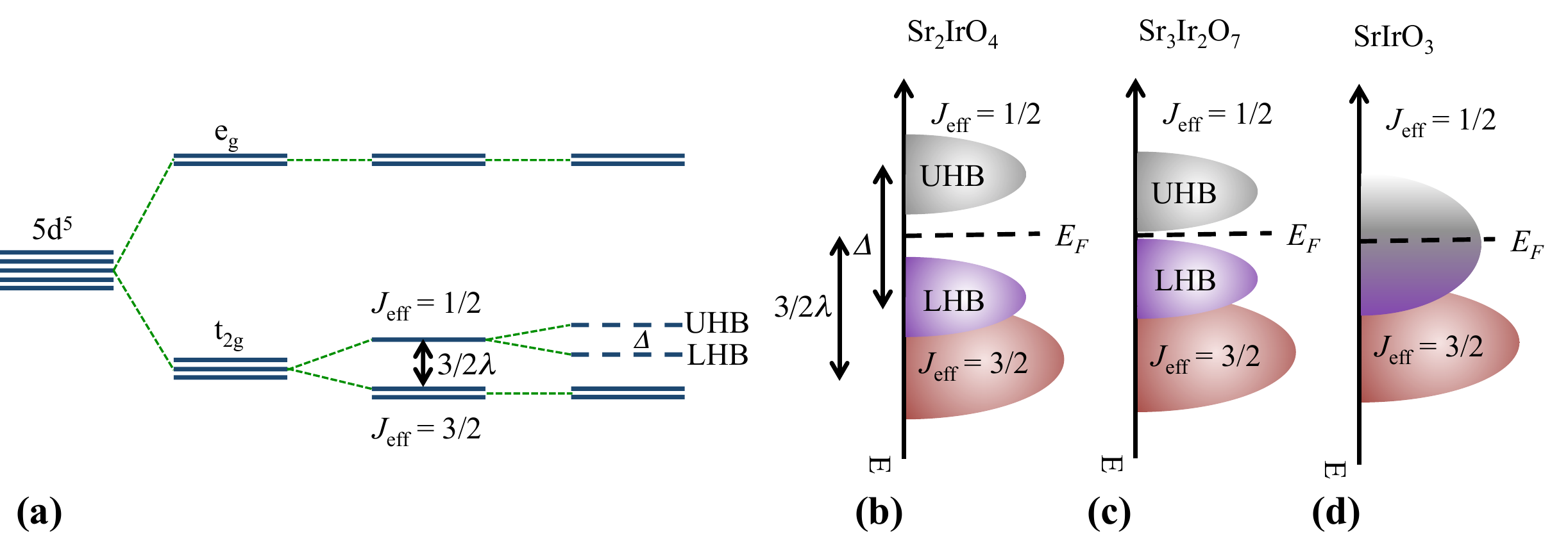}
\caption{(a) Crystal-field splitting, spin-orbit coupling, $\lambda$, and Coulomb induced Mott-gap, $\Delta$, in Sr$_{n+1}$Ir$_{n}$O$_{3n+1}$. (b)-(d) Schematic electronic structure of Sr$_{n+1}$Ir$_{n}$O$_{3n+1}$.}
\label{fig:CFandMG}
\end{figure*}

The first reported magnetic \gls*{tr}-\gls*{RIXS} experiments were conducted on Sr$_{2}$IrO$_{4}$ \cite{Dean2016Ultrafast}, a model quasi-two-dimensional square-lattice quantum antiferromagnet \cite{jackeli2009mott,Wang_prl_2011,Kim_prl_2012}. Long-range magnetic order occurs below $T_{N}\approx 240$~K and arises from the interplay of several degrees of freedom \cite{Gao_prb_1998, Kimprl2008}. Each Ir atom has a $5d^5$ electronic configuration and sits at the center of an oxygen octahedron. All five $d$-electrons reside in the t$_{2g}$ level due to the large crystal field splitting. Strong spin-orbit coupling $\sim$400~meV further splits the t$_{2g}$ level into the $J_\text{eff} = 3/2$ state occupied by 4-$5d$ electrons, and a single-occupied $J_\text{eff} = 1/2$ state. The Ir atoms form a square net resembling that of the copper ions in the layered high-T$_C$ cuprates, and a modest Coulomb repulsion induces an insulating band gap of 600~meV (see Fig.~\ref{fig:CFandMG}a and b). Sr$_{2}$IrO$_{4}$ belongs to the Ruddlesden-Popper series of layered iridates Sr$_{n+1}$Ir$_n$O$_{3n+1}$. The band-gap of the spin-orbit Mott insulating state successively reduces with increasing neighboring Ir-O layers, $n$. In the $n\to\infty$ end member SrIrO$_{3}$ the gap is closed making the system metallic (c.f.\ Fig.~\ref{fig:CFandMG}c and d) \cite{Hao2017On}.

\begin{figure*}[tbh]
\centering
\includegraphics[width=0.9\linewidth]{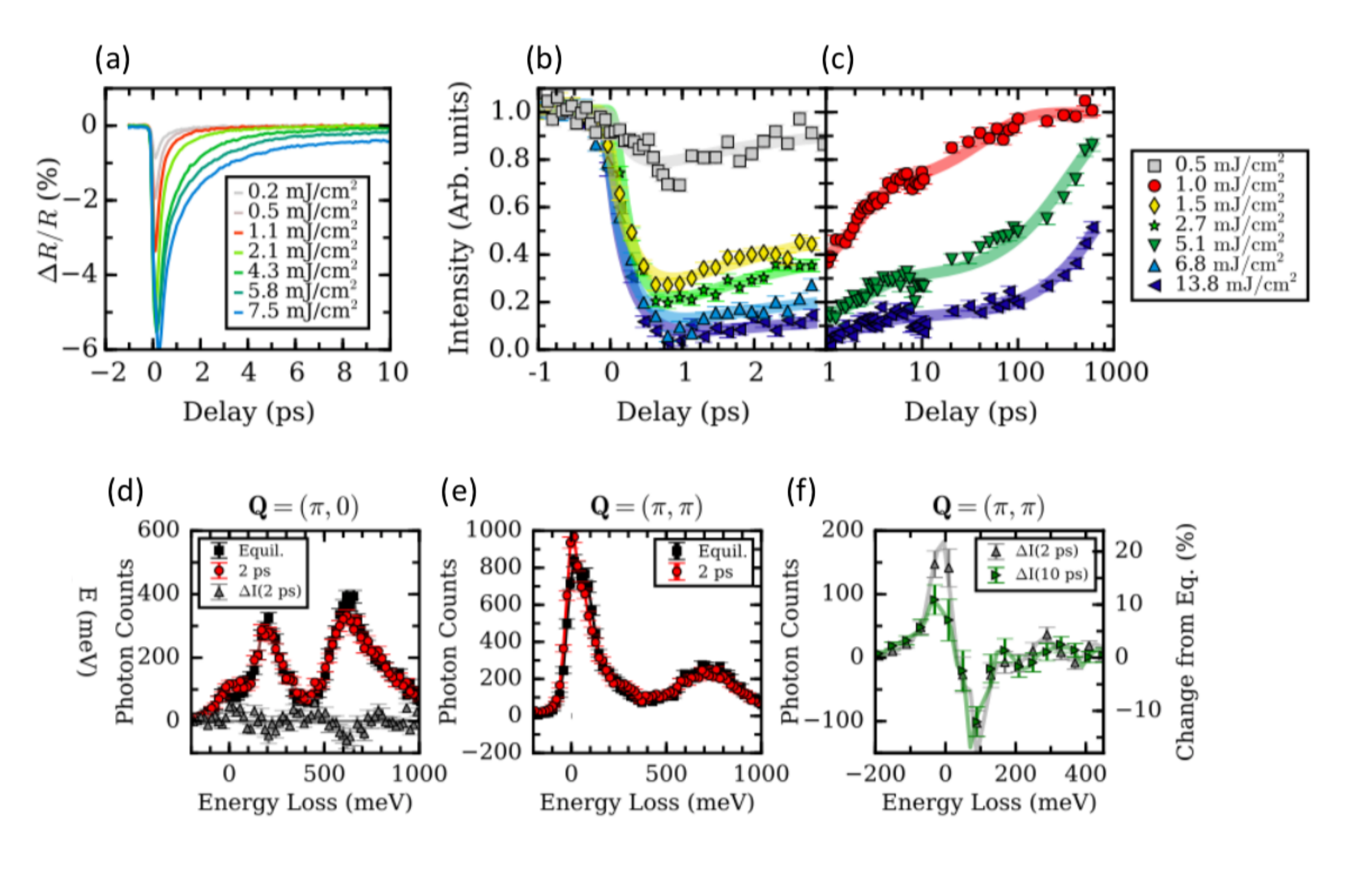}
\caption{Time dependence of (a) the relative optical reflectivity ($\Delta R / R$) and (b) and (c) the magnetic Bragg peak in Sr$_{2}$IrO$_{4}$ after 620~meV photo-excitation at different laser fluences. Excitation spectrum at (d) $\textbf{Q} = (\pi, 0)$ and (e) $\textbf{Q} = (\pi, \pi)$ in the non-perturbed state and 2~ps after the laser pulse, under the pump fluence of 6~mJ/cm$^2$. The difference in the RIXS spectrum before and after the optical excitation at $\textbf{Q} = (\pi, \pi)$ is shown in (f). Taken from Ref.~\cite{Dean2016Ultrafast}.}
\label{fig:spectra}
\end{figure*}

The charge degree of freedom in these materials was studied by time-resolved optical reflectivity measurements \cite{Dean2016Ultrafast}. Here electrons are excited from the lower (LHB) into the upper Hubbard band (UHB) by 2~$\mu$m laser pulses with an energy that matches the electronic band gap. Figure~\ref{fig:spectra}(a) shows the relevant timescales in Sr$_{2}$IrO$_{4}$ that are needed to recombine the charge carriers under different laser fluences. The plot reveals a decay time that is attributed to the excitation of the charge carriers, and a fast and slow recovery process in the sub-picosecond and few-picosecond regime, respectively.

Since the magnitude of the electron bandwidth, the Coulomb repulsion and the spin-orbit coupling share similar energy scales in the iridates, the same 2~$\mu$m (620~meV) laser pulse also affects the magnetic properties \cite{Dean2016Ultrafast}. A combined \gls*{REXS} and \gls*{RIXS} study at the hard X-ray Ir $L_{3}$ edge probed the energy, momentum and time-dependent response of the transient magnetic state at $T = 110$~K (c.f.\ Fig.~\ref{fig:rixs_process}). Figure~\ref{fig:spectra}(b) and (c) display the time and fluence dependence of a magnetic Bragg peak in Sr$_{2}$IrO$_{4}$. After the initial suppression of peak intensity following the optical excitation, the three-dimensional long-range antiferromagnetic order fully recovers within several hundred picoseconds. Intriguingly, a partial restoration of magnetism is observed already within a few picoseconds. This is further clarified by tr-\gls*{RIXS}, as discussed below.

\begin{figure*}[tbh]
\includegraphics[width=\linewidth]{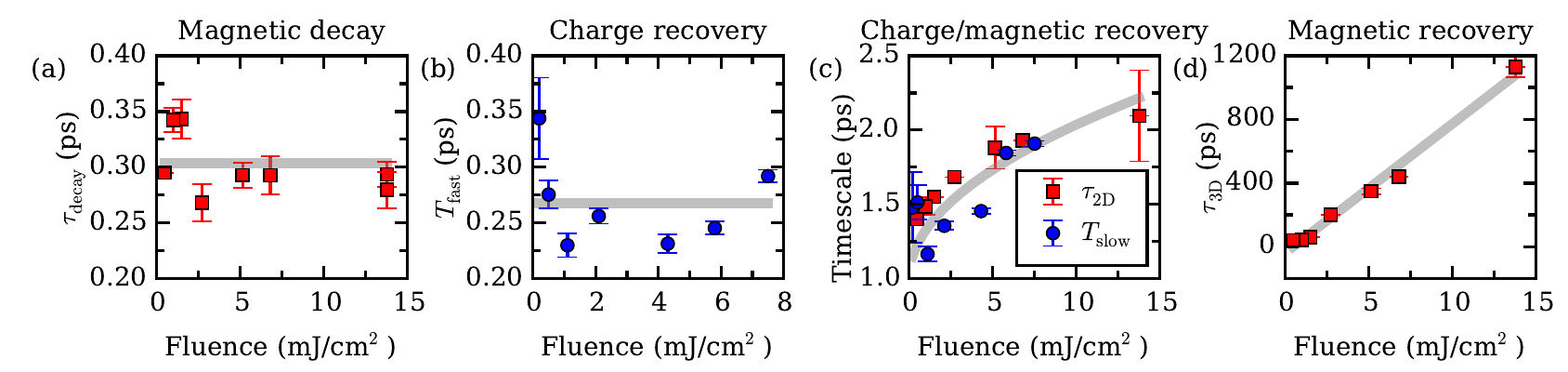}
\caption{(a) Fluence dependence of the decay time that is required to destroy the ground state. (b) Fast recovery of the charge degree of freedom measured via optical reflectivity. (c) Slow recovery of the charge degree of freedom that matches the recovery timescale of the 2D magnetic correlations. (d) Slow recovery of the 3D magnetic long-range order. Taken from \cite{Dean2016Ultrafast}.}
\label{fig:time_evolution}
\end{figure*}

The \gls*{RIXS} spectra displayed in Fig.~\ref{fig:spectra}(d-f) show the first ever view of magnetic short-range correlations within a photo-excited ultrafast transient state. Magnetic excitations were measured before, and 2~ps after, the arrival of pump laser pulses, and with a pump fluence large enough to fully suppress the 3D magnetic order \cite{Dean2016Ultrafast}. The low-energy excitation spectrum of Sr$_{2}$IrO$_{4}$ features a dispersing spin-wave below 200~meV (see Fig.~\ref{fig:spectra}(d)) and an orbital excitation of the $J_{\text{eff}} = 1/2$ state around 600~meV. The main result of the study demonstrates that despite the destruction of the long-range magnetic order, magnons are already observed 2~ps after the impact of the laser pulse. Furthermore, the recovery timescale of these predominantly two-dimensional in-plane magnetic fluctuations matches the partial recovery of long-range magnetic order as found by time-resolved \gls*{REXS} (shown in Fig.~\ref{fig:time_evolution} and discussed below).

The various timescales and their fluence dependences are shown in Fig.~\ref{fig:time_evolution}. Both charge and magnetic degrees of freedom exhibit fast and slow recovery dynamics \cite{Dean2016Ultrafast}. Most strikingly, the recovery timescale of the 2D magnetic fluctuations matches that of slower charge recovery, providing direct evidence for a coupling between them. The authors further suggest that the slower restoration of 3D magnetic order may be attributed to incoherently oriented IrO$_{2}$ planes along the tetragonal axis.

This pioneering experiment paves the way for further time-resolved \gls*{REXS}/\gls*{RIXS} experiments that may lead to a full microscopic understanding of the correlated ground state in the  Sr$_{n+1}$Ir$_{n}$O$_{3n+1}$ series. While electrons were pumped across the Mott-gap, the bandwidth, Coulomb repulsion and spin-orbit coupling were not tuned directly in this work. A selective pump of the Ir-O bonds in Sr$_{2}$IrO$_{4}$ with mid-infrared ultrafast laser pulses, for instance, could tune the crystal-field environment and modulate the magnetic interaction strength of the system\footnote{In general, infrared pulses are not the only ones used in a time-resolved XFEL experiments. A plethora of fundamental collective excitations between 0.3$\sim$3~THz can be optically targeted using optically-pumped organic crystals \cite{kubacka2014}.}. Observing the time-dependent evolution of the transient state after shaking the Ir-O bonds will not only disentangle the lattice and electronic degrees of freedom, but may also demonstrate how to drive the material into specific magnetic states by pumping another degree of freedom (in this case, the lattice). The role of Coulomb repulsion may be clarified by investigating other members in the series. For example, Sr$_{3}$Ir$_{2}$O$_{7}$ features two closely-coupled Ir-O layers in the crystal structure and shows major modifications in the materials properties \cite{Biswas_intech_2016,Kim_prl_2012_2}. This material lies on the verge of a metal-to-insulator transition with a heavily reduced Mott-gap (see Fig. \ref{fig:CFandMG} c) \cite{Biswas_intech_2016}. Magnetic long-range order emerges below $T_{N}\approx 285$~K with a magnetic moment orientation perpendicular to the tetragonal basal plane \cite{Boseggia_prb_2012}. The excitation spectrum strongly deviates from the isotropic Heisenberg model that describes Sr$_{2}$IrO$_{4}$ with a much-increased magnetic gap. Thus, the restoring forces for the magnetic ground state in Sr$_{3}$Ir$_{2}$O$_{7}$ are expected to be different from those in Sr$_{2}$IrO$_{4}$, and a study of their dynamics following a pump may further clarify the underlying processes. 

In contrast to other members of the series, SrIrO$_{3}$ is a correlated paramagnetic metal close to the metal-to-insulator transition (see Fig. \ref{fig:CFandMG} d) \cite{Biswas_intech_2016}. It can be grown as thin films on substrates where the relative lattice mismatch between the film and the substrate provides an additional tuning parameter. The epitaxial strain induces changes in the Ir-O bond angle and could drive the system over the metal-to-insulator transition. Thus, the variation of film thickness and substrate material enables a selective realization of the material properties. This further adds to the possibilities of altering the magnetic ground state by optical means, and may finally allow one to fully disentangle the coupled degrees of freedom in the series.

The strong experimental push towards the understanding of electronic short-range correlations in transient states parallels several recent theoretical studies of the Hubbard model that is believed to describe the main properties of many strongly correlated materials \cite{Eckstein2016ultra,Werner2012,Bittner2018EDMFT,Kogoj2014polaron,Du2017thermalization,Wang2017coherent,Tsuji2013PRL,Mentinknatcommun2015,Mentinkrevew2017,Secchi2013AnnalsPhysics}. Particularly relevant in view of the tr-RIXS study on Sr$_{2}$IrO$_{4}$ are the results in Ref. \cite{Eckstein2016ultra} that report the interaction of short-range spin fluctuations with the charge degree of freedom in photo-excited Mott insulators. Using non-equilibrium dynamical mean-field theory, the study provides evidence for short-range magnons in a two-dimensional paramagnetic system with relaxation times in the femtosecond time domain. The extension of the theoretical model to an antiferromagnetically ordered state may lead to a microscopic understanding of the time difference between 2D fluctuations and 3D order in Sr$_{2}$IrO$_{4}$ \cite{Dean2016Ultrafast}.

\section{Vision for the future}\label{sec:future}
As demonstrated in previous sections, \gls*{tr}-\gls*{RIXS} has unique advantages for investigating quantum magnetism in transient states. Here we discuss future developments of \gls*{tr}-\gls*{RIXS} instrumentations and the scientific problems that will be addressed because of these advances.

\subsection{New XFELs and RIXS spectrometers}
Hitherto, progress in \gls*{tr}-\gls*{RIXS} has been slowed by the limited availability of suitable X-ray sources and spectrometers. In the past 10 or so years since the first \gls*{XFEL}s came online, their primary focus has been on numerous exciting opportunities in more tractable techniques such as \gls*{XAS} and X-ray diffraction. Efforts to realize more complex experiments are, however, already well under way. A major increase in the availability and quality of \gls*{XFEL} sources is likely to drive further progress in realizing highly challenging experiments such as \gls*{tr}-\gls*{RIXS}. In particular the more established sources such as the Hamburg \gls*{FLASH}, the Trieste \gls*{FERMI}, the Stanford \gls*{LCLS} and the \gls*{SACLA} will be joined by the European \gls*{XFEL}, \gls*{PAL-XFEL}, the Stanford LCLS-II and \gls*{SwissFEL} \cite{flash2006, elettra2012, lcls2010, sacla2012,europeanXFEL2017,pal2017, swissfel2010}. Dramatic improvements in brightness, stability, pulse length and pulse repetition rate are expected. The small \gls*{RIXS} cross-section may particularly benefit from improved time-average spectral brightness that can potentially exceed that of synchrotrons significantly, making feasible new experiments e.g. probing magnetism in single-layer films. These improvements are particularly beneficial when ultra-high energy resolution is pursued.  

The spectrometer required to analyze the X-ray energy also represents a significant challenge, but several teams have successfully performed experiments at \gls*{XFEL}s \cite{Rusydi2014,dell2016extreme,Dean2016Ultrafast,Mitrano2018Ultrafast}. To date, \gls*{XFEL}-based \gls*{RIXS} tends to lag behind synchrotron experiments in terms of energy and momentum resolution. Many of the technical developments implemented at synchrotrons \cite{dvorak2016towards, Brookes2018beamline} can potentially be employed at \gls*{XFEL}s provided further practical complications, such as limited experimental floor space, compatibility with other experiments, finite beam stability, can be avoided or overcome. In many cases shorter spectrometers ($ \lessapprox 2$~m compared to $\sim 15$~m at synchrotrons) are more practical at \gls*{XFEL}s. A recent compact soft X-ray spectrometer design has delivered down to 300~meV resolution at 1~keV incident energy \cite{chuang2017modular, Mitrano2018Ultrafast}. Hard X-ray experiments with a 1~m spectrometer achieved $70$~meV at 11.2~keV \cite{Dean2016Ultrafast}. Looking to the future, multiplexing techniques that can efficiently measure the \gls*{RIXS} incident energy dependence \cite{Chuang2016Multiplexed}, streaking approaches to measuring the time-delay or analyzing the scattered x-ray polarization \cite{braicovich2014simultaneous} will be interesting routes to consider.

\subsection{Evolution of tr-RIXS}
Driven by these technical developments we anticipate considerable advances in \gls*{tr}-\gls*{RIXS} in the coming years. A primary focus will be far more detailed and precise characterizations of spin, charge and orbital behaviors in transient states, from which opens routes disentangling the complex interplay between these different degrees of freedom. For example, time-resolved diffraction measurements have recently quantified how terahertz-frequency optical pulses modify the crystal structure of cuprates \cite{mankowsky2014nonlinear}. \gls*{RIXS} could potentially be used to quantify how magnetic exchange is modified in such a transient state. One can also consider using \gls*{RIXS} to evaluate change in spin stripe-magnon interactions in similar transient states \cite{Miao2017Charge}. These types of coupling are generic to many different types of correlated system including $3d$, $4d$ and $5d$-electron oxides such as cuprates, nickelates and osmates -- systems in which steady-state \gls*{RIXS} is becoming more insightful \cite{dean2013persistence, Fabbris2017Doping, Calder2016spin}. By carefully choosing the optical pump, it would be desirable to selectively drive the different forms of order in the material. With the help of RIXS, it is then possible to address the chicken and egg interdependencies of these order parameters. More ambitiously, the application of higher laser pump fluences, and oscillatory Fluoquet-type pluses can potentially access transient states not adiabatically connected to the states at thermal equilibrium. The extent to which these states have novel properties remains to be seen \cite{Wang2018theoretical}. It is also notable that \gls*{FT}-\gls*{IXS} has recently observed the decay of optical phonons into pairs of acoustic phonons \cite{Teitelbaum2018Direct}, direct measurement of similar processes between spin waves would be very insightful in conceptualization the evolution of magnetic states out of equilibrium.

%We can also use RIXS to interrogate the relationship between charge, spin and orbital degrees of freedom. For example, in systems with coupled structural and orbital transitions, we could address the chicken and egg interdependences of these order parameters by studying whether laser excitation can create states with only one form of order. A particularly challenging area of ultrafast science aims to create new types of order which are not adiabatically connected to the ground state, e.g. driving a paramagnetic material into a magnetically ordered state. Since these states will have a finite lifetime the magnetic correlation length of the order parameter might be short -- so the high sensitivity of \gls*{RIXS} may come into play to detect the weak diffuse signals expected in this class of experiments.

\subsection{New types of RIXS at XFELs}
The high brilliance of new sources should allow `diffract-then-destroy' types of \gls*{RIXS} experiment for detecting the collective excitations in quantum materials under extreme conditions at equilibrium. These involve problems in which the state of interest only exists fleetingly, but the experiment can be performed within the short X-ray pulse duration before the state is destroyed. The most prominent example of this approach is in protein crystallography where diffraction data is collected before the sample is destroyed by radiation damage. With expected higher \gls*{XFEL} peak brightness, similar ideas will also likely become important for ultrafast quantum materials research. Single-shot \gls*{XFEL} experiments in pulsed magnetic fields have been demonstrated using X-ray diffraction, paving the way for investigations of magnetic excitation under large magnetic fields in in quantum materials \cite{Gerber2015}. One can also envisage accessing ultra-high pressures transiently, following themes in \gls*{XFEL}-based geology research and other areas. An as yet unexploited idea would be to try to measure \gls*{RIXS} at extremely low temperatures, where the sample temperature would be elevated by beam heating under normal circumstances. 

In standard \gls*{IXS} experiments, one determines the dynamical properties of the sample in the frequency domain through the energy dependence of $S^{\alpha \beta}({\bm Q}, \omega)$. \gls*{XFEL}s open the opportunity to determine $S^{\alpha \beta}({\bm Q}, \omega)$ in the time domain. This approach has been illustrated in recent experiments by Trigo and collaborators \cite{Trigo2013}. In this study, the structural X-ray diffuse scattering of crystalline germanium was measured as a function of time delay after photo-excitation. Fourier transforming the oscillatory component of the diffuse scattering yielded the low-energy phonon dispersion. In the limit of negligible pump intensity, the phonon energy dispersion obtained should be the same as that accessible by standard \gls*{IXS}, leading to this technique being dubbed \gls*{FT}-\gls*{IXS}. Notably, the energy resolution of \gls*{FT}-\gls*{IXS} is determined not by the spectrometer as in conventional \gls*{IXS}, but by the longest delay possible, so \gls*{FT}-\gls*{IXS} has good sensitivity to slow, low-energy excitations. Clever choices of photo-excitations may also open way to picking out modes of particular interest, which have small structure factors in standard \gls*{IXS} measurements. Combining \gls*{FT}-\gls*{IXS} with atomic core-hole resonance may allow access to electronic excitations from the charge, spin and orbital degrees of freedom (see Sec.~\ref{sec:rixs}) similar to \gls*{RIXS}. This would represent what might be thought of as Fourier-transform or time-domain \gls*{RIXS}, but as far as we are aware such experiments have not yet been successfully performed.  Some of these low-energy charge and orbital excitations, e.g.\ amplitudons, phasons, orbitons, etc.\ are not accessible by neutrons, and could require energy resolution well below that of any spectrometer-based RIXS at XFELs and synchrotrons either today or anticipated over the next decade. Experimentally identifying these charge/orbital excitations and understanding their evolution in the temperature-doping phase diagram would be particularly interesting for cuprates and Fe-based superconductors, especially since these excitations are theoretically proposed as critical to the formation of superconductivity \cite{fradkin2015colloquium, lee2009ferro}.

Not only the time structure, but also the large peak intensity of \gls*{XFEL} pulses will allow access to novel types of RIXS processes. To date, in the vast majority of \gls*{RIXS} experiments, the X-ray has been assumed to be a linear perturbation to the material \cite{ament2011resonant}. Under this assumption, the RIXS spectrum is interpreted as originating from the interaction of a single photon with the sample, where the incident photon density is low enough for there to be a negligible probability of one photon encountering the effects of other photons. \gls*{XFEL}s now deliver peak photon intensity sufficient to access non-linear processes. In general two cases are distinguished, in which the stimulating X-rays arise either from a separate X-ray beam, or they arrive from scattered photons in the sample. The former case is often referred to as stimulated Raman scattering and has only been realized using photons in the ultra violet regime so far \cite{Ferrori2016}. The latter case takes advantage of the finite energy distribution in the \gls*{XFEL} beam, leading to an amplified spontaneous emission that emerges from noise around the center of the X-ray pulse \cite{Rohringer2012}. The last few years have seen $2p$ core hole stimulated emission in crystalline silicon proving the feasibility of solid-state non-linear experiments \cite{Beye2013stimulated}. As emphasized by Beye and collaborators \cite{Beye2013stimulated} stimulated processes can enhance the efficiency of \gls*{RIXS} by several orders of magnitude and confine emission into a well-defined cone allowing the scattered photons to be more efficiently collected. Non-linear \gls*{RIXS} therefore has potential to circumvent radiation damage problems. Perhaps even more interestingly, one can use the enhanced specificity of such process to access,  for example, hidden order parameters and excitations, such as those only accessible via quadrapole resonances \cite{Wang2017On}.

\section{Summary}
Recent years have seen the advent of \gls*{tr}-\gls*{RIXS} as a flexible and rich probe of non-equilibrium phenomena. We argue that this technique has great potential for measuring spin, charge, orbital and lattice excitations after photo-excitation as a means of unpicking how these degrees of freedom intertwine to form emergent states and for characterizing the new laser-driven transient states. The increasing number of operating \gls*{XFEL} facilities and \gls*{tr}-\gls*{RIXS} instruments will not only serve as a basis to solve long-standing scientific questions, but also to investigate physics beyond the limit of linear response theory and to enable the development of novel techniques such as stimulated or Fourier-transform \gls*{RIXS}. Ultra-bright X-ray pulses may also be used to perform future experiments on samples susceptible to radiation damage or at extremely low temperatures and high magnetic fields, accessing phase-space regions that could not be reached via X-ray scattering before. Thus, the prospected capabilities of tr-RIXS in its various forms are pointing towards a bright future.

\section{Acknowledgements}
We acknowledge D.~Zhu, R.~Mankowsky, V.~Thampy, X.~M.~Chen, J.~G.~Vale, D.~Casa, Jungho Kim, A.~H. Said, P.~Juhas, R.~Alonso-Mori, J.~M.~Glownia, A.~Robert, J.~Robinson, M.~Sikorski, S.~Song, M.~Kozina, H.~Lemke, L.~Patthey, S.~Owada, T.~Katayama, M.~Yabashi, Yoshikazu Tanaka, T.~Togashi, Jian Liu, C.~Rayan Serrao, B.~J.~Kim, L.~Huber, C.-L.~Chang, D.~F.~McMorrow, and M.~F\"{o}rst for the considerable joint effort in demonstrating \gls*{tr}-\gls*{RIXS}. We thank J.~St\"{o}hr and G.~Ingold for valuable discussions.  This work is supported by the U.S.\ Department of Energy, Office of Basic Energy Sciences, Early Career Award Program under Award No.\ 1047478. Work at Brookhaven National Laboratory was supported by the U.S. Department of Energy, Office of Science, Office of Basic Energy Sciences, under Contract No.\ DE-SC0012704. The work at Argonne National Laboratory was supported by the U.S.\ Department of Energy, Office of Basic Energy Sciences, under Contract No.\ DE-AC0206CH11357. The work at ShanghaiTech U.\ was partially supported by MOST of China under the grand No.\ 2016YFA0401000. The work at ICFO received financial support from Spanish MINECO (Severo Ochoa grant SEV-2015-0522), Ram\'on y Cajal programme RYC-2013-14838, FIS2015-67898-P (MINECO/FEDER), Fundaci\'o Privada Cellex, and CERCA Programme / Generalitat de Catalunya. D.G.M.\ acknowledges funding from the Swiss National Science Foundation, Fellowship No.\ P2EZP2\_175092.

\bibliographystyle{rsta}
\bibliography{refs}

\end{document}